\documentclass[12pt]{article}

\usepackage{ieee6x9}
\usepackage{amsmath,amssymb,graphics,amsthm,psfrag,subfigure}
\usepackage{amsfonts,epsfig,graphicx,subfigure}
\usepackage{amsmath,amssymb,amsthm}
\usepackage{epic,eepic}
\usepackage{epsf}
\usepackage{fancyheadings}
\usepackage{graphics}
\usepackage{amsfonts}
\usepackage{amsmath}
\usepackage{amssymb}
\usepackage{epic}
\usepackage{eepic}
\usepackage{eepicemu}
\usepackage{psfrag}

\newlength{\cheatLength}
\input{arxivmacros.sty}

\makeatletter
\long\def\@makecaption#1#2{
        \vskip 0.8ex
        \setbox\@tempboxa\hbox{\small {\bf #1:} #2}
        \parindent 1.5em  
        \dimen0=\hsize
        \advance\dimen0 by -3em
        \ifdim \wd\@tempboxa >\dimen0
                \hbox to \hsize{
                        \parindent 0em
                        \hfil 
                        \parbox{\dimen0}{\def\baselinestretch{0.96}\small
                                {\bf #1.} #2
                                } 
                        \hfil}
        \else \hbox to \hsize{\hfil \box\@tempboxa \hfil}
        \fi
        }
\makeatother

\renewcommand{\thefootnote}{} 

\begin{document}

\title{Low density codes achieve the rate-distortion bound}
\footnotetext{Emin Martinian was supported by Mitsubishi Electric Research Labs while Martin Wainwright was supported by NSF Grant 
DMS-0528488 and an Alfred P. Sloan Foundation Fellowship.}

\author{
\begin{tabular}{cc}
\normalsize Emin Martinian & \normalsize Martin Wainwright \\
%
\normalsize Mitsubishi Electric Research Labs & \normalsize University
of California at Berkeley 
\\ 
\normalsize Cambridge, MA 02139 & \normalsize Berkeley, CA 94720\\
\normalsize Email: \{martinian\}@merl.com & \normalsize Email:
\{wainwrig\}@eecs.berkeley.edu 
\end{tabular}
}

\date{}


\maketitle

\vspace*{-2.8in}%
\vbox to 2.8in{\small\tt%
\begin{center}
  \begin{tabular}[t]{cl}
   Published in: &  Proceedings of the Data Compression Compression, \\
&  Snowbird, Utah.  March 2006
  \end{tabular} 
\end{center}
\vfil}

\thispagestyle{empty}
\pagestyle{empty}
\renewcommand{\thefootnote}{\arabic{footnote}} 

\normalsize

{\bf Abstract:} We propose a new construction for low-density source
codes with multiple parameters that can be tuned to optimize the
performance of the code.  In addition, we introduce a set of analysis
techniques for deriving upper bounds for the expected distortion of
our construction, as well as more general low-density constructions.
We show that (with an optimal encoding algorithm) our codes achieve
the rate-distortion bound for a binary symmetric source and Hamming
distortion.  Our methods also provide rigorous upper bounds on the
minimum distortion achievable by previously proposed low-density
constructions.

\section{Introduction}
\vspace{\cheatLength}

While low-density parity check (LDPC) codes can provably approach the
channel coding capacity~\cite{Richardson:it:2001} for point-to-point
transmission, currently there are relatively few theoretical results
on low-density codes for lossy source coding, channel coding with
encoder side information, and source coding with decoder side
information.  Note that the latter three scenarios all involve some
aspect of quantization.  Even though quantization and error correction
are closely related, the standard LDPC constructions used for channel
coding generally fail~\cite{martinian:2003:allerton}.  One viable
option is trellis-based quantization (TCQ)~\cite{Marcellin90}, which
has been used both for lossy source coding, as well as for distributed
source coding~\cite{Wang01,Chou03,Yang05}.  However, saturating
fundamental bounds with TCQ requires taking the constraint length to
infinity~\cite{Viterbi74}, which incurs exponential complexity even
for message-passing decoders/encoders.  Consequently, it is of
considerable interest to develop low-density constructions that are
also capable of saturating the information-theoretic bounds.

Previous work~\cite{martinian:2003:allerton} has shown that
low-density generator matrix (LDGM) codes, which are dual to LDPC
codes, are provably optimal for binary erasure quantization (a special
type of source coding). This motivates the use of LDGM codes and
variants for more general compression problems.  Indeed, recent
work~\cite{Ciliberti05b,Murayama04,wainwright:2005:isit} has shown
empirically that LDGM codes, in conjunction with variants of
sum-product message-passing for encoding, can approach the
rate-distortion bound for a binary symmetric source (BSS).  In
addition, non-rigorous replica or cavity method
calculations~\cite{Ciliberti05b,Murayama04} also suggest that the
theoretical performance of LDGM codes is close to optimal.

This paper makes two primary contributions to this area.  First, we
propose a new low-density construction for lossy source coding with
multiple parameters that can be tuned to optimize the performance of
the code.  Our construction includes as a special case the ordinary
LDGM codes examined
previously~\cite{Ciliberti05b,martinian:2003:allerton,Murayama04,wainwright:2005:isit}.
Second, we develop methods useful for analyzing the expected
distortion of our constructions as well as more general lossy source
codes.  Using these methods, we show that (with optimal quantization)
our codes saturate the rate-distortion bound for a uniform binary
source and Hamming distortion.  As we will show in a longer version of
this paper, our methods also lead to rigorous upper bounds on the
distortion achievable by a standard LDGM construction. Thus, provided
that a low complexity iterative encoding algorithm can be found, our
results suggest that low density codes can provide significant
improvements for a wide class of quantization problems.

The remainder of this paper is organized as follows.  After
introducing some notation, we describe our new low density generator
matrix construction in \secref{sec:comp-constr} and bound its
performance in \secref{sec:main-results}.  In particular, we describe
the tools required to analyze low density source codes through a
series of lemmas, which we believe illustrate the main insights of the
paper.  Finally, we close with some concluding remarks in
\secref{sec:concluding-remarks} and postpone all proofs to the appendix.


\paragraph{{\bf{Notation:}}} 
Vectors/sequences are denoted in bold (\eg, $\sSrc$), random variables
in sans serif font (\eg, $\rvSrc$), and random vectors/sequences in
bold sans serif (\eg, $\rvsSrc$).  Similarly, matrixes are denoted
using bold capital letters (\eg, $\genMat$) and random matrixes with
bold sans serif capitals (\eg, $\rvGenMat$).  We use $I(\cdot;\cdot)$,
$H(\cdot)$, and $\rent{\cdot}{\cdot}$ to denote mutual information,
entropy, and relative entropy (Kullback-Leibler distance),
respectively.  Finally, we use $\cardinal{\{\cdot\}}$ to denote the
cardinality of a set, $\pNorm{\cdot}{p}$ to denote the $p$-norm of a
vector, and $\binent{t}$ to denote the entropy of a Bernoulli($t$)
random variable.

\section{The Compound Construction}
\label{sec:comp-constr}
\vspace{\cheatLength}

\newcommand{\ratetop}{\ensuremath{R_{\ensuremath{t}}}}
\newcommand{\ratebot}{\ensuremath{R_{\ensuremath{b}}}}
\newcommand{\midbit}{\ensuremath{m}}
\newcommand{\rateover}{\ensuremath{R}}

\begin{figure}[h]
\begin{center}
\psfrag{#k#}{$k$} \psfrag{#m#}{$\midbit$} \psfrag{#topdeg#}{$\topdeg$}
\psfrag{#vdeg#}{$\vdeg$} \psfrag{#cdeg#}{$\cdeg$} \psfrag{#n#}{$n$}
\widgraph{.8\textwidth}{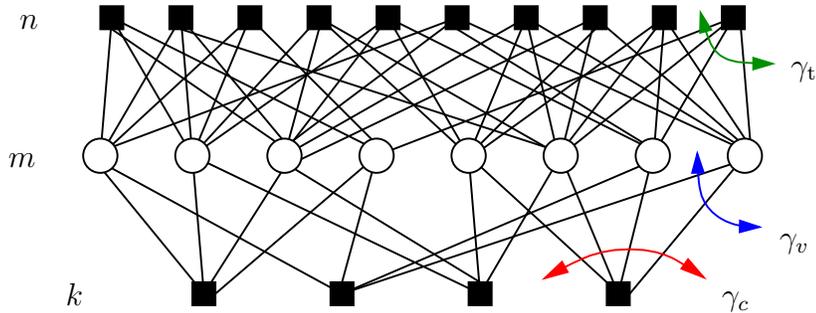}
\caption{Illustration of the compound code construction, involving an
LDGM (top section) with $\topdeg = 4$ and an LDPC (bottom section)
with $(\vdeg, \cdeg) = (2,4)$. 
\label{fig:compound-construction}}
\end{center}
\end{figure}

The construction considered in the paper is illustrated
in~\figref{fig:compound-construction}: the top section consists of an
LDGM code $\topCode$ of rate $\ratetop = \frac{\midbit}{n}$ with $n$
source bits and $\midbit$ information bits, whereas the bottom section
consists of an LDPC code of rate $\ratebot = 1 - \frac{k}{\midbit}$
with $\midbit$ bits constrained by $k$ checks.  The compound code
formed by joining the top and bottom code can generate $2^{\ratebot
\midbit} = 2^{\midbit - k}$ possible source reconstructions of length
$n$, so that the overall code $\fullCode$ has rate $\rateover =
\ratetop \, \ratebot$.
Note that a check-regular LDGM code corresponds to the special case of
setting $\ratebot = 1$.

To quantize a length $n$ binary source vector $\sSrc$ using the
compound construction, an encoder finds an assignment for the
$\midbit$ bits in the middle layer that satisfy the constraints of the
bottom LDPC code.  Formally, we can denote the $\midbit$-by-$n$
generator matrix for the top LDGM code as $\genMat$ and the
$k$-by-$\midbit$ parity check matrix for the bottom LDPC code as
$\parMat$.  Then $\sQuant$ is a codeword of the overall code if
$\sQuant = \sIntm \, \genMat$ and $\parMat \, \sIntm' = 0$ for
some assignment of the middle layer, which we denote as $\sIntm$.
Thus, an optimal encoder for $\sSrc$ would find the codeword 
minimizing the Hamming distance, $\hDist{\sIntm\genMat}{\sSrc}$, such
that $\parMat\sIntm'=0$.  Since the vector $\sIntm$ has length
$\midbit$, storing or transmitting $\sIntm$ directly would achieve
only compression rate $\ratetop$.  Instead, we can use the fact that
there are only $2^{\midbit - k}$ valid choices for $\sIntm$, to store
$\sIntm$ using only $k$ bits, resulting in compression rate
$\rateover$.  For example, we could store the $k$-bit information
vector that when encoded with the bottom LDPC code yields $\sIntm$.

\mypara{Random LDPC Ensemble:} For the bottom LDPC code, we use the
standard $(\vdeg, \cdeg)$-regular LDPC ensemble studied by
Gallager \cite{gallager:ldpc:book}.  Specifically, each of the
$\midbit$ variable nodes in the 
middle layer connects to $\vdeg$ check nodes in the bottom layer.
Similarly, each of the $k$ check nodes in the bottom layer connects to
$\cdeg$ variable nodes in the middle layer.  For convenience, we
restrict ourselves to even check degrees $\cdeg$.  Note that these
degrees are linked to the rate via the relation $\frac{\vdeg}{\cdeg} =
1 -\ratebot$.  A random LDPC code $\cldpc \equiv \cldpc(\vdeg, \cdeg)$
is generated by choosing uniformly from this ensemble. \\

\vspace*{.01in}


\mypara{Random LDGM ensemble:} For the top LDGM code, each of the
$\nbit$ checks at the top are randomly connected to $\topdeg$ variable
nodes in the middle layer chosen uniformly at random.  This leads to a
Poisson degree distribution on the information bits and makes the
resulting distribution of a random codeword easy to characterize:
\begin{lems}
\label{LemInducedDmin}
Let $\rvGenMat$ be a random generator matrix obtained by placing
$\topdeg$ ones in each column uniformly at random.  Then for any
vector $\sIntm \in \{0,1\}^\midbit$ with a fraction of $\weight$ ones,
the distribution of the corresponding codeword $\sIntm \,
\rvGenMat$ is Bernoulli($\inducedWeight{\weight;\topdeg}$) where
\begin{equation}
\inducedWeight{\weight;\topdeg} = \frac{1}{2} \cdot \left[1 - (1-2
  \weight)^{\topdeg}\right].
\end{equation}
\end{lems}

\section{Main Results}
\label{sec:main-results}
\vspace{\cheatLength}

Although our methods apply to the compound construction more
generally, we state our main result in application to the special case
with $\ratetop = 1$ and $\ratebot = \rate$.  For these choices, we can
guarantee that our compound construction approaches the optimal
rate-distortion trade-off as the blocklength $\nbit$ tends to infinity
using finite choices of degrees in our LDGM/LDPC
construction.\footnote{Our methods also yield upper bounds on the
achievable distortion of the check-regular LDGM construction ($\ratetop =
\rate$ and $\ratebot = 1$).  Subsequent work will describe the use of
alternative rate pairs $(\ratetop, \ratetop)$ for source and channel
coding with side information.}

\btheos
\label{ThmMain}
Consider an arbitrary rate distortion pair $(\distor,
\rate(\distor))$.  For any $\gap > 0$. there exists a finite LDGM
degree $\topdeg(\gap, \distor)$ and an LDPC code with finite degrees
$\vdeg(\gap, \distor)$ and $\cdeg(\gap, \distor)$ such that a randomly
chosen code with rate $\rate \defeq \rate(\distor) + \gap$ in the
associated LDGM ensemble achieves distortion $\distor$ with
probability $1-\exp (-c n)$ for some constant $c$.
\etheos
\noindent 
As a particular example of our results illustrated later in
\figref{fig:gap}, the degree choices $\topdeg = 4$, $\botdeg = 8$ for
rate $\rate(\distor) = 1/2$, are sufficient to make the gap $\Delta$ zero (within the precision of our numerical calculations).
The proof of \thrmref{ThmMain} consists of several steps, which we motivate
and describe in the following text.  Proofs of these auxiliary results
are provided in the appendix.

\subsection{Expected Number of Good Codewords}

For a length $\blockLength$ code $\fullCode$ and a source vector
$\sSrc$, we define $\goodWords{\fullCode}{\sSrc}{D}$ to be the number
of codewords that are within Hamming distance $D\blockLength$ of
$\sSrc$.  Specifically, let $\indic{i}{\fullCode}{\sSrc}{D}$ be 1 if
the $i$th codeword in the compound code $\fullCode$ is within Hamming
distance $D\blockLength$ of the source $\sSrc$, and 0 otherwise.  Then
\begin{equation}
\goodWords{\fullCode}{\sSrc}{D} \defeq \sum_i \indic{i}{\fullCode}{\sSrc}{D}.
\end{equation}

Ideally, $\goodWords{\fullCode}{\sSrc}{D}$ should be large and there
should be many good codewords provided that the rate exceeds the
rate-distortion function: $\rate > 1-\binEnt{D}$.  Specifically, if we
consider a random source vector $\rvsSrc$ and a randomly generated
code $\rvFullCode$, then the probability that the code is successful
is simply\footnote{When the source and/or code are random, then we
drop the indexing of random quantities and write
$\indic{i}{\rvFullCode}{\rvsSrc}{D}$ and
$\goodWords{\fullCode}{\sSrc}{D}$ as random variables $\rvIndic{i}{D}$
and $\rvGoodWords{D}$.}  $\Pr[ \rvGoodWords{D} > 0 ]$.

Since analyzing this probability directly is generally difficult, most
random coding arguments \cite{cover91:book} consider the expectation
$E[ \rvGoodWords{D} ]$.  For essentially any code (and in particular
the compound construction), it is possible to show that the expected
number of good codewords is large:
\blems
\label{lem:expValLowerBound}
\begin{equation}
E[ \rvGoodWords{D} ]\geq \frac{1}{\nbit+1} 2^{\blockLength \big( \rate
- \big[1 - \ent(\distor) \big] \big)}.
\end{equation}
\elems

\subsection{Typical Number of Good Codewords}

Unfortunately, the fact that the expected number of codewords is large
is insufficient to show that the code achieves the rate-distortion
bound.  Rather, in order to show that the code is good, we must show
that the typical number of good codewords is not too far from the
expected number of good codewords (or at least non-zero).  For high
density codes, this can be done by using Chebyshev's inequality, which
depends on the variance of $\rvGoodWords{D}$.  For most low density
constructions (including our own), the variance is too large for
Chebyshev's inequality to yield a useful bound.  Consequently, we
instead use Shepp's second moment method as summarized in the
following proposition:\footnote{\propref{prop:second-moment-method}
can be established by defining an indicator random variable,
$\rvOneGoodWord{D}$, for the event $\{\rvGoodWords{D} > 0\}$ and
applying the Cauchy-Schwartz inequality to obtain
$\E[\rvGoodWords{D}]^2 = \E[\rvGoodWords{D}\rvOneGoodWord{D}]^2 \leq
\E[\rvGoodWords{D}^2]\cdot \E[\rvOneGoodWord{D}]$, which is equivalent
to the desired result.}
\begin{prop}
\label{prop:second-moment-method}
For any positive integer valued random variable $\rvz$,
$\Pr[\rvz > 0] \geq \frac{\E[\rvz]^2}{\E[\rvz^2]}$.
\end{prop}

To show that there is typically at least one good codeword, we
must upper bound $\E[\rvGoodWords{D}^2]$, which can be cast in a more
useful form using the following lemma:
\begin{lems}
\label{lem:second-moment-simplification}
\begin{equation}
\label{eq:second-moment-simplification}
\E[\rvGoodWords{D}^2] = E[\rvGoodWords{D}] + E[\rvGoodWords{D}] \cdot
\left\{ \sum_{j \neq 0} \Pr[ \rvIndic{j}{D}=1\mid\rvIndic{0}{D}=1]\right\}\\
\end{equation}
\end{lems}

\lemref{lem:second-moment-simplification} illustrates one of the main
differences between low density and high density constructions.
Specifically, in a high density construction, each codeword can be
chosen independently yielding $\Pr[
\rvIndic{j}{D}=1\mid\rvIndic{0}{D}=1] = \Pr[ \rvIndic{j}{D}=1]$ and
implying $\E[\rvGoodWords{D}^2] \leq E[\rvGoodWords{D}] +
E[\rvGoodWords{D}]^2$.  In contrast, for low density codes, there will
usually be some dependence between the codewords.  For example, in the
usual LDGM construction, when the information bits $\sIntm$ have low
weight, then the resulting codeword $\sIntm\,\genMat$ will also
have low weight.  Consequently, if the all-zero codeword is within
Hamming distance $D\blockLength$ of the source, then these low weight
codewords probably are as well and so $\Pr[
\rvIndic{j}{D}=1\mid\rvIndic{0}{D}=1]$ can be much larger than $\Pr[
\rvIndic{j}{D}=1]$.  In particular, we can bound $\Pr[
\rvIndic{j}{D}=1\mid\rvIndic{0}{D}=1]$ by considering the weight of
the information sequence $\sIntmC{j}$ used to generate the $j$th codeword:
\begin{lems}
\label{lem:LemUpperBern}
Let $\sIntmC{j}\,\rvGenMat$ be the $j$th codeword obtained by
multiplying a weight $\weightC{j}$ vector $\sIntmC{j}$ by a random
matrix from the LDGM ensemble and let $\rvIndic{j}{D}$ denote the
event that codeword $j$ is within Hamming distance $D\blockLength$ of
a random Bernoulli($1/2$) source.  Then for any even degree $\topdeg$,
letting $\weightC{0}=0$ yields
\begin{equation}
\label{eq:EqnUpperBern}
\Pr[\rvIndic{j}{D}=1\mid\rvIndic{0}{D}=1]  \leq  
\begin{cases}  1 & 
\mbox{if $0 \leq \weightC{j} \leq \keyupper(\distor; \topdeg)$} \\
 2^{- \blockLength \kull{\distor}{ \inducedDmin{\weightC{j}}}} &
\mbox{otherwise},
\end{cases}
\end{equation}
where 
\begin{equation}
\label{EqnKeyUpper}
\keyupper(\distor; \topdeg) = \frac{1}{2} \biggr[1 - \big(1 - 2
\distor \big)^{\frac{1}{\topdeg}} \biggr].
\end{equation}
\end{lems}

\lemref{lem:LemUpperBern} shows that
$\Pr[\rvIndic{j}{D}=1\mid\rvIndic{0}{D}=1]$ is small whenever the
weight of the information sequence for a codeword is large.  So to
characterize the sum over this probability we must consider how many
vectors of a given weight in the middle layer satisfy the constraints
of the bottom LDPC code $\botCode$.  Specifically, we denote the average (log
domain) weight enumerator of $\botCode$ (\ie, the rate of codewords of
$\botCode$ with a given weight) as
\begin{equation}
\label{eq:weight-enumerator-def}
\WtEnumCodeWt{\botCode}{\omega} \defeq \frac{1}{\blockLength} \cardinal{
  \{\rvsQuant \in \botCode \mid \pNorm{\rvsQuant}{1} = \omega\cdot\blockLength
  \}}. \ 
\end{equation}
Intuitively, by combining (\ref{eq:weight-enumerator-def}) with
\lemref{lem:LemUpperBern}, we 
can bound the term in braces of (\ref{eq:second-moment-simplification}):
\begin{equation}
\sum_{j \neq 0} \Pr[
  \rvIndic{j}{D}=1\mid\rvIndic{0}{D}=1]\leq
  \sum_{t=1}^{\keyupper(\distor; \topdeg)}
  2^{\WtEnumCodeWt{\botCode}{t/\blockLength}} 
  + \sum_{t=\keyupper(\distor; \topdeg)}^{\blockLength}
  2^{\blockLength \left[\WtEnumCodeWt{\botCode}{t/\blockLength} -
  \kull{\distor}{ \inducedDmin{t}}\right]}.
\end{equation}
Formally, we can use this idea to obtain the following result:
\begin{theos}
\label{th:gap}
Consider a sequence of rate $\rate$ compound codes of increasing
blocklength $\blockLength$.  Suppose that the following inequality
holds for all sufficiently large blocklengths:
\begin{equation}
\label{eq:gap}
\rate - [1 - \binEnt{D}] > \frac{1}{\blockLength} \log \left\{
\sum_{t=1}^{\keyupper(\distor; \topdeg)}
2^{\WtEnumCodeWt{\botCode}{\frac{t}{\blockLength}}} +
\sum_{t=\keyupper(\distor; \topdeg)}^{\blockLength} 2^{\blockLength
\left[\WtEnumCodeWt{\botCode}{\frac{t}{\blockLength}} -
\kull{\distor}{ \inducedDmin{\frac{t}{\blockLength}}}\right]} \right\}
\end{equation}
Then the probability that a code in the sequence fails to quantize a
source with distortion at most $D\blockLength$ goes to zero as
$\blockLength\rightarrow\infty$.
\end{theos}

\subsection{Reducing Dependency Between Codewords}

\begin{figure}[h]
\begin{center}
\psfrag{YLABEL}{$\frac{1}{n} \log (\cdot)$}
\widgraph{.7\textwidth}{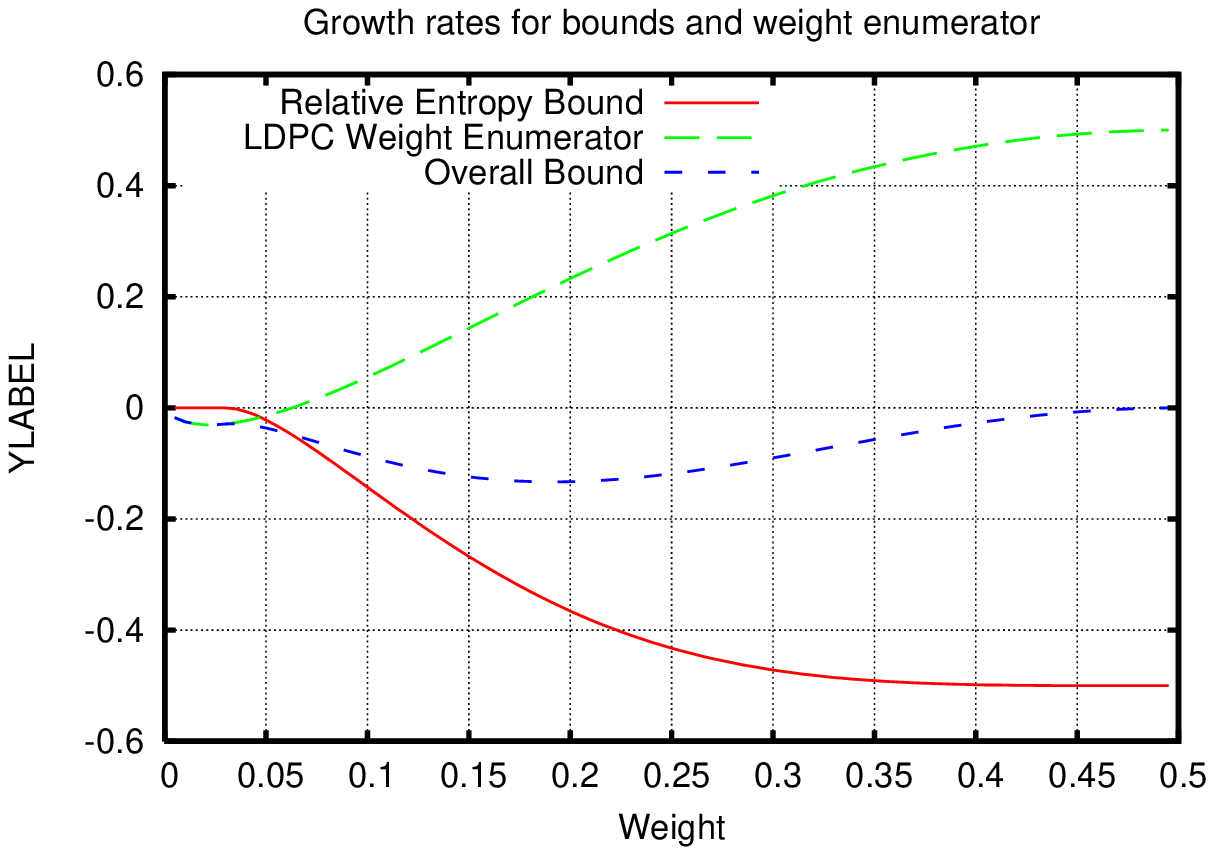}
\caption{Log of bounds and weight enumerator for $\rate=1/2$,
  $\topdeg=4$, $\botdeg=8$, at distortion $D\approx 0.11$ normalized
  by the blocklength $\blockLength$.  The relative entropy bound from
  (\ref{eq:EqnUpperBern}) is zero for weights below
  $\keyupper(\distor;\topdeg)$ and then quickly goes to
  $2^{-\blockLength/2}$.  The (log domain) weight enumerator for a
  regular rate $1/2$ LDPC code is negative for weights below the
  minimum distance and then rises to $2^{\blockLength/2}$.  As long as
  the relative entropy bound is stronger than the weight enumerator,
  the excess rate in (\ref{eq:gap}) of \thrmref{th:gap} will be
  negligible.
\label{fig:gap}}
\end{center}
\end{figure}

The bracketed term on the RHS of (\ref{eq:gap}) corresponds to the
{\em excess rate} required beyond the minimum $1-\binEnt{D}$ and is plotted
in \figref{fig:gap} for the compound code in
\figref{fig:compound-construction}.  The first term represents the number
of low weight codewords of the bottom code.  Since the bound from
\lemref{lem:LemUpperBern} does not become active until weight
$\keyupper(\distor; \topdeg)$, making the first term negligible
requires choosing the LDPC ensemble so that the minimum distance is
greater than the weight $\keyupper(\distor; \topdeg)$ resulting from
the choice of the degree $\topdeg$ in the LDGM ensemble.  The exponent
of the second term in (\ref{eq:gap}) is the sum of the weight
enumerator and the bound from \lemref{lem:LemUpperBern}.  For this
term to be negligible, the bottom LDPC code must have a weight
enumerator that grows less quickly than the error exponent in
(\ref{eq:EqnUpperBern}).

Using the exact formula for the asymptotic weight enumerator of
regular LDPC codes developed by Litsyn and Shevelev
\cite{Litsyn_and_Shevelev}, it is possible to prove the following
result:\footnote{The proof essentially requires showing that the sum
  of the weight enumerator and the bound from
  \lemref{lem:LemUpperBern} is negative for all weights in
  $[\keyupper(\distor;\topdeg),1/2]$.  This can be done by checking the
  appropriate derivatives of the sum and is omitted for brevity.}
\begin{prop}
\label{prop:painful-stuff}
There exist choices for $\topdeg$, $\vdeg$, and $\cdeg$ such that the
term in braces in (\ref{eq:gap}) becomes negligible.
\end{prop}

\section{Concluding Remarks}
\label{sec:concluding-remarks}
\vspace{\cheatLength}

In this paper, we proposed a new construction for low density source
codes and introduced tools to analyze low density generator matrix
codes.  As stated in Lemma~\ref{lem:LemUpperBern} and illustrated in
\figref{fig:gap}, our main insight was that the source coding
performance of a low density code can be bounded by considering the
weight of the codewords.  Thus, by using a compound code to control
the weight spectrum we obtained codes that approach the
rate-distortion function.  A future paper will describe and analyze
these types of compound constructions in application to source and
channel coding with side information.

\appendix

\section{Proofs}
\vspace{\cheatLength}

\begin{proof}[Proof of \lemref{LemInducedDmin}:]
By construction of the LDGM ensemble for $\rvGenMat$, each bit of the
codeword $\sIntm \, \rvGenMat$ is independent of the others and is
the modulo-2 sum of $\topdeg$ randomly and independently selected bits
of $\sIntm$.  So the resulting codeword has a Bernoulli distribution
and all that remains is to determine the probability that a given bit
is one, which we denote as $\inducedDmin{\weight}$.

For any output bit, let the random variable $\rvValueForBit{i}$ denote
whether the $i$th one in a column of $\rvGenMat$ occurs in a position
where $\sIntm$ has a one (\ie, $\rvValueForBit{i}$ is the value of the
variable node connected to the $i$th link of a given check node at the
top of \figref{fig:compound-construction}).  Then
$\inducedDmin{\weight}$ is exactly the probability that $\sum_{i=1}^{\topdeg}
\rvValueForBit{i}$ is even.  Letting $\zTransInducedDmin{\weight}{z}$
denote the generating function (\ie, the $z$-transform) of $\sum_{i=1}^{\topdeg}
\rvValueForBit{i}$ yields
\begin{align}
\label{eq:ztrans-start}
\inducedDmin{\weight}-&(1-\inducedDmin{\weight}) =
\zTransInducedDmin{\weight}{z=-1} = \prod_{i=1}^{\topdeg}
\left(\Pr[\rvValueForBit{i}=0] +
  z^{-1}\Pr[\rvValueForBit{i}=1]\right)\bigg|_{z=-1}\\
\label{eq:ztrans-result}
&= (1-\weight +\weight\cdot z^{-1})^{\topdeg}\bigg|_{z=-1} =
(1-2\weight)^{\topdeg}.
\end{align}
Equating the leftmost term of (\ref{eq:ztrans-start}) and the
rightmost term of (\ref{eq:ztrans-result}) and solving for
$\inducedDmin{\weight}$ yields the desired result.
\end{proof}

\begin{proof}[Proof of \lemref{lem:expValLowerBound}:]
\begin{align}
E[\rvGoodWords{D}] &= E\left[\sum_i \rvIndic{i}{D}\right] = \sum_i
E[\rvIndic{i}{D}] = \sum_i \Pr[ \hDist{\rvsQuantC{i}}{\rvsSrc} \leq D
\blockLength] \\  
& \geq \sum_i \frac{2^{-\blockLength
    \kull{D}{1/2}}}{(\blockLength+1)^2} = \sum_i
\frac{2^{-\blockLength [1 - \binEnt{D}]}}{(\blockLength+1)^2} = 
\frac{2^{\blockLength \{R - [1 - \binEnt{D}]\}}}{(\blockLength+1)^2} 
\end{align}
The first line follows by repeatedly expanding the definition of the
random variables $\rvGoodWords{D}$ and $\rvIndic{i}{D}$.  For the
next line, we lower bound the probability that a given codeword
$\rvsQuantC{i}$ is within distortion $D\blockLength$ using standard
large deviations results (Theorem 12.1.4, \cite{cover91:book}).  Note
that nothing in this argument depends on the actual code construction
itself (except for the number of codewords).
\end{proof}

\begin{proof}[Proof of \lemref{lem:second-moment-simplification}:]
\begin{align}
\label{eq:thing-to-upper-bound}
\E[&\rvGoodWords{D}^2] = \E\left[\sum_{i} \sum_{j}
  \rvIndic{i}{D}\rvIndic{j}{D}\right] = E[\rvGoodWords{D}] +
\sum_{i} \sum_{j \neq i} \E\left[ \rvIndic{i}{D}\rvIndic{j}{D}\right]
\\
%
%
\label{eq:use-def-of-rvIndic}
&= E[\rvGoodWords{D}] +
\sum_{\sSrc} \sum_{i} \sum_{j \neq i} \Pr[
\hDist{\rvsQuantC{j}}{\sSrc}\leq D\blockLength,
\hDist{\rvsQuantC{i}}{\sSrc}\leq D\blockLength]\Pr[\rvsSrc=\sSrc] \\
\label{eq:add-codeword-i-to-both-size}
&= E[\rvGoodWords{D}] +
\sum_{\sSrc} \sum_{i} \sum_{j \neq i} \Pr[
\hDist{\rvsQuantC{j}\oplus\rvsQuantC{i}}{\sSrc\oplus\rvsQuantC{i}}\leq
D\blockLength, \notag\\
&\hspace{2in} \hDist{0}{\sSrc\oplus\rvsQuantC{i}}\leq
D\blockLength ]\Pr[\rvsSrc=\sSrc] \\
\label{eq:bring-in-primes}
&= E[\rvGoodWords{D}] +
\sum_{\sSrc} \sum_{i} \sum_{j'\neq 0} \Pr[
\hDist{\rvsQuantC{j'}}{0}\leq
D\blockLength, \hDist{0}{\sSrc'}\leq
D\blockLength ]\Pr[\rvsSrc=\sSrc] \\
\label{eq:back-to-probs}
&= E[\rvGoodWords{D}] +
\sum_{i} \sum_{j'\neq 0} \Pr[\rvIndic{j'}{D}=1, \rvIndic{0}{D}=0]\\
\label{eq:separate-probs}
&= E[\rvGoodWords{D}] +
\left\{\sum_{i}\Pr[\rvIndic{0}{D}=1]\right\}\cdot\left\{ \sum_{j'\neq 0}
  \Pr[\rvIndic{j'}{D}=1| \rvIndic{0}{D}=1]\right\}
\end{align}
To obtain (\ref{eq:thing-to-upper-bound}) we consider the diagonal
terms separately from the off-diagonal terms, and note that
$E[\rvIndic{i}{D}^2] = E[\rvIndic{i}{D}]$ since the $\rvIndic{i}{D}$
are indicator variables.  Next, we apply the definition of
$\rvIndic{i}{D}$ to get (\ref{eq:use-def-of-rvIndic}) and then add
$\rvsQuantC{i}$ to each side of the $\hDist{\cdot}{\cdot}$ terms to
obtain (\ref{eq:add-codeword-i-to-both-size}).  Since the code is
linear, adding the codewords $\rvsQuantC{i}$ and $\rvsQuantC{j}$
yields another codeword which we denote $\rvsQuantC{j'}$.  This
observation combined with writing $\sSrc' =
\sSrc\oplus\rvsQuantC{i}$ yields (\ref{eq:bring-in-primes}).  To go
from (\ref{eq:bring-in-primes}) to (\ref{eq:back-to-probs}), we note
that for a uniformly random source, $\Pr[\rvsSrc=\sSrc] =
\Pr[\rvsSrc=\sSrc']$.  Finally, to obtain the desired result from
(\ref{eq:separate-probs}), we note that $\rvIndic{i}{D}$ is
independent of $i$ and hence $E[\rvIndic{i}{D}] = E[\rvIndic{0}{D}]$.
\end{proof}

\begin{proof}[Proof of \lemref{lem:LemUpperBern}:]
  We focus on the case when $\inducedDmin{\weightC{j}} \geq \distor$.
  Solving this relation for $\weightC{j}$ yields the formula for
  $\keyupper(\distor;\topdeg)$ in (\ref{EqnKeyUpper}) and so
  $\inducedDmin{\weightC{j}} < \distor$ corresponds to the trivial
  bound in the top case of (\ref{eq:EqnUpperBern}).  Thus, for
  $\inducedDmin{\weightC{j}} \geq \distor$ we have
\begin{align}
\label{eq:just-expand-def-of-rvIndic}
&\hspace{-2pt}\Pr\left[\rvIndic{j}{D}=1\right. |\
\rvIndic{0}{D}=1\left.\right] \leq 
\Pr\left[\hDist{\rvsQuantC{j}}{\rvsSrc}\leq D\blockLength\ |\ 
\hDist{\rvsSrc}{0} \leq D\blockLength\right]\\
&\hspace{-2pt}\stackrel{(a)}{\leq} \max_{t \leq \distor\blockLength}
\Pr\bigg[\hDist{\rvsQuantC{j}}{1^{t}0^{\blockLength - t}}
\leq \distor\blockLength\bigg]
\stackrel{(b)}{\leq}\Pr\bigg[\hDist{\rvsQuantC{j}}{0^{\blockLength}}
\leq \distor\blockLength\bigg]
\stackrel{(c)}{\leq}
2^{-\blockLength\kull{\distor}{\inducedDmin{\weightC{j}}}}.
\label{eq:use-Sanov-for-binary-vars}
\end{align}
We obtain (\ref{eq:just-expand-def-of-rvIndic}) from the definition of
the random variable $\rvIndic{j}{D}$.  For (a), since
$\rvsQuantC{j}$ is a Bernoulli sequence, without loss of generality we
can imagine that all the ones in $\rvsSrc$ occur at the start of the
sequence.  To obtain an upper bound, we put in as many such ones
as required to maximize the desired probability.  In (b), we note that
$\inducedDmin{\weightC{j}} \leq 1/2$ implies that it is more likely
that a given position of 
$\rvsQuantC{j}$ is zero than one so $t=0$ gives the largest value
for the maximization.  Finally, to obtain (c), 
we apply Sanov's Theorem (Theorem 12.1.4, \cite{cover91:book}).  Note that the
reason we required $\inducedDmin{\weightC{j}} \geq \distor$ originally
is that this condition is required by Sanov's Theorem in (c).
\end{proof}

Before proving \thrmref{th:gap}, we require the following lemma:
\begin{lems}
\label{lem:polynomial-small-error}
  For a compound code that satisfies
  (\ref{eq:gap}), $\Pr[\rvGoodWords{D} > 0] > (1/2)\cdot(\blockLength+1)^{-2}$.
\end{lems}
\begin{proof}
First, assume that $\sum_{j\neq
  0}\Pr[\rvIndic{j}{D}\mid\rvIndic{0}{D}]\geq 1$ because if this is
not the case then (\ref{eq:second-moment-simplification}) immediately
implies that $\Pr[\rvGoodWords{D} > 0] \geq 1/2$ and the proof is
complete.  Therefore continuing from the assumption that $\sum_{j\neq
  0}\Pr[\rvIndic{j}{D}\mid\rvIndic{0}{D}]\geq 1$ yields
\begin{align}
\label{eq:minor-lemma-start-eq}
\Pr[\rvGoodWords{D} > 0] &\stackrel{(a)}{\geq}
\frac{\E[\rvGoodWords{D}]^2}{\E[\rvGoodWords{D}^2]} \stackrel{(b)}{=}
\frac{\E[\rvGoodWords{D}]^2}{\E[\rvGoodWords{D}] \{1 +
  \sum_{j\neq 0} \Pr[\rvIndic{j}{D}\mid\rvIndic{0}{D}]\}}\\
\label{eq:minor-lemma-middle-eq}
&\stackrel{(c)}{\geq}
\frac{\E[\rvGoodWords{D}]^2}{2\cdot\E[\rvGoodWords{D}] \cdot \{
  \sum_{j\neq 0} \Pr[\rvIndic{j}{D}|\rvIndic{0}{D}]\}}
=
\frac{\E[\rvGoodWords{D}]/2}{
  \sum_{j\neq 0} \Pr[\rvIndic{j}{D}|\rvIndic{0}{D}]}\\
&\stackrel{(d)}{\geq}
\frac{\E[\rvGoodWords{D}]/2}{2^{\blockLength\{\rate-[1-\binEnt{D}]\}}}
\stackrel{(e)}{\geq}
\frac{2^{\blockLength\{\rate-[1-\binEnt{D}]\}}}{2(\blockLength+1)^2\cdot
  2^{\blockLength\{\rate-[1-\binEnt{D}]\}}} = \frac{1}{2(\blockLength+1)^2}
\end{align}
where (a) follows
from \propref{prop:second-moment-method}, 
(b) comes from
\lemref{lem:second-moment-simplification}, (c) follows from the
assumption in the first sentence, (d) comes from
(\ref{eq:gap}), and (e) comes from \lemref{lem:expValLowerBound}. 
\end{proof}

\begin{proof}[Proof of \thrmref{th:gap}:]
\lemref{lem:polynomial-small-error} tells us that the 
  probability that at least codeword is found within distortion $D$ is
  at at least $(1/2)/(\blockLength+1)^2$, \ie, there is at least a
  small chance that a good codeword exists.  The key insight of the
  remainder of the proof is that while $(1/2)/(\blockLength+1)^2$ may
  be small, it is not {\em exponentially} small.  Hence if we can show
  that the distortion for a compound code is concentrated near its
  typical value except with some {\em exponentially} small
  probability, then \lemref{lem:polynomial-small-error} immediately
  implies that the event $\{\rvGoodWords{D} > 0\}$ must correspond to
  the typical distortion.  To prove exponential concentration, we
  show that the actual error probability, $\Pr[\rvGoodWords{D}=0]$ is
  smaller than $e^{-c \blockLength}$ for some constant $c$ using 
  martingale arguments \cite{Alon:book:2000,Richardson:it:2001}.  

  Specifically, we define a Doob martingale $\martingale{i}{\botCode}$
  that is the expected value of the distortion between the best
  codeword and the source (conditioned on the bottom code $\botCode$)
  when the first $i$ columns of the generator matrix $\rvGenMat$ (\ie,
  the connections from the first $i$ checks to their respective
  variables) of the top code in 
  \figref{fig:compound-construction} have been revealed.  Going
  from step $i$ to $i+1$ and revealing check $i+1$ can
  only change the value of the martingale $\martingale{i}{\botCode}$
  by at most 1.  Hence, by the Azuma-Hoeffding inequality, the
  probability that a sample path of the martingale differs from its
  expected value by more than $\epsilon$ is less than $2
  e^{-\blockLength \epsilon^2}$.

  Since \lemref{lem:polynomial-small-error} shows that the probability
  that $\Pr[\rvGoodWords{D}>0]$ is at least an inverse polynomial (and 
  hence {\em not} exponentially small), the event
  $\{\rvGoodWords{D}>0\}$ must determine the expected value of the
  martingale.  Therefore other events that result in a distortion
  larger than $D$ (\eg, $\{\rvGoodWords{D}=0\}$) must be exponentially
  small. 
\end{proof}

\begin{proof}[Proof of \thrmref{ThmMain}:]
Combining \thrmref{th:gap} with \propref{prop:painful-stuff}
establishes this result.
\end{proof}

\vspace{-.4in}
\small
\bibliographystyle{latex8}
\bibliography{refs}

\end{document}